\documentclass[11pt]{article}

\usepackage[margin=25mm]{geometry}

\usepackage{setspace}
\newcommand{\pytorch}{PyTorch}
\newcommand{\jax}{JAX}
\newcommand{\tensorflow}{TensorFlow}

\usepackage{booktabs}

\usepackage{blindtext}

\usepackage{amsmath}
\usepackage{amssymb}

\newcommand{\Rbb}{\mathbb{R}}
\newcommand{\Cbb}{\mathbb{C}}

\newcommand{\Conj}[1]{ { \text{conj}\left(#1\right) } }

\usepackage{mathtools}

\usepackage[numbers]{natbib}
\usepackage{titlesec}
\titleformat{\section}{\centering\bf \sffamily }{\thesection.}{0.5em}{}
\titleformat{\subsection}{\centering\bf \sffamily }{\thesubsection.}{0.5em}{}

\usepackage{framed}
\usepackage{xcolor}
\colorlet{shadecolor}{gray!20!white}

\usepackage{pythonhighlight}
\usepackage{listings}

\usepackage{xcolor}

\definecolor{codegreen}{rgb}{0,0.6,0}
\definecolor{codegray}{rgb}{0.5,0.5,0.5}
\definecolor{codepurple}{rgb}{0.58,0,0.82}
\definecolor{backcolour}{rgb}{0.95,0.95,0.92}

\lstdefinestyle{mystyle}{
    backgroundcolor=\color{backcolour},   
    commentstyle=\color{codegreen},
    keywordstyle=\color{magenta},
    numberstyle=\tiny\color{codegray},
    stringstyle=\color{codepurple},
    basicstyle=\ttfamily\footnotesize,
    breakatwhitespace=false,         
    breaklines=true,                 
    captionpos=b,                    
    keepspaces=true,                 
    numbers=left,                    
    numbersep=5pt,                  
    showspaces=false,                
    showstringspaces=false,
    showtabs=false,                  
    tabsize=2
}

\usepackage{hyperref}

\usepackage{microtype}

\usepackage{cleveref}

\creflabelformat{equation}{#2#1#3}

\usepackage{amsthm}

\renewcommand{\paragraph}[1]{ { \sffamily \textbf{#1} } }

\begin{document}

\title{\Large%
	\sffamily 
	\textbf{A tutorial on automatic differentiation with complex numbers}
}	
\author{%
	{Nicholas Krämer}%
	\footnote{Contact: \texttt{pekra(at)dtu.dk}.}
	\\
	\\
	Technical University of Denmark
	\\
	Kongens Lyngby, Denmark
}

\maketitle

\begin{abstract}
\noindent
Automatic differentiation is everywhere, but there exists only minimal documentation of how it works in complex arithmetic beyond stating ``derivatives in $\Cbb^d$'' $\cong$ ``derivatives in $\Rbb^{2d}$'' and, at best, shallow references to Wirtinger calculus.
Unfortunately, the equivalence $\Cbb^d \cong \Rbb^{2d}$ becomes insufficient as soon as we need to derive custom gradient rules, e.g., to avoid differentiating ``through'' expensive linear algebra functions or differential equation simulators.
To combat such a lack of documentation, this article surveys forward- and reverse-mode automatic differentiation with complex numbers, covering topics such as Wirtinger derivatives, a modified chain rule, and different gradient conventions while explicitly avoiding holomorphicity and the Cauchy--Riemann equations (which would be far too restrictive).
To be precise, we will derive, explain, and implement a complex version of Jacobian-vector and vector-Jacobian products almost entirely with linear algebra without relying on complex analysis or differential geometry.
This tutorial is a call to action, for users and developers alike, to take complex values seriously when implementing custom gradient propagation rules -- the manuscript explains how. 
\end{abstract}

\section{Introduction}

\paragraph{Differentiable programming.}
A computer program is a sequence of function evaluations.
If each function is differentiable, and if we know each derivative, the chain rule combines the individual derivatives in the derivative of the computer program.
In recent years, this automatic differentiation has been crucial for gradient-based optimisation of sophisticated machine learning models through libraries like \pytorch{} \citep{ansel2024pytorch}, \tensorflow{} \citep{abadi2015tensorflow}, autograd \citep{maclaurin2015autograd}, \jax{} \citep{jax2018github}, or programming languages like Julia \citep{bezanson2017julia}, or Dex \citep{maclaurin2019dex}.
However, the impact of automatic differentiation far exceeds machine learning applications.
For example, works like that by \citet{abdelhafez2019gradient}, \citet{luchnikov2021riemannian}, or \citet{tamayo2018automatic} use automatic differentiation for quantum computations in physics and chemistry, which places a high demand on complex arithmetic, perhaps unlike traditional machine learning applications like natural language processing or computer vision.
And even though \citet[page 51]{maclaurin2016modeling} explains how real-valued automatic differentiation need not be changed much to extend to complex arithmetic, applying techniques like implicit differentiation \citep{blondel2022efficient} or adjoint sensitivity analysis \citep[e.g.][]{cao2003adjoint} to complex values is all but obvious to most practitioners.
Unfortunately, this lack of transparency affects performance: custom gradient rules via, for instance, implicit differentiation and adjoint sensitivity analysis are critical for efficient gradients of structured solution routines, for example, those in numerical linear algebra \citep[e.g.][]{kraemer2024gradients,roberts2020qr}.

\paragraph{Contributions and outline.}
These notes are a tutorial, and everything they contain is known in some form or another, even though some derivations might be new.
\Cref{section-related-work} discusses related work, but it can be skipped on the first read because the subsequent sections don't depend on it.
Our main goal is to assemble Jacobian-vector and vector-Jacobian products for functions with complex in- and outputs.
To get there, \Cref{section-background-automatic-differentiation} recalls background on real-valued automatic differentiation, and \Cref{section-complex-autodiff-as-real-autodiff} extends it to $\Cbb$ via $\Cbb \cong \Rbb^2$.
However, \Cref{section-complex-autodiff-as-real-autodiff} also shows how automatic differentiation via $\Cbb \cong \Rbb^2$ can be needlessly laborious, which is why these notes don't stop at \Cref{section-complex-autodiff-as-real-autodiff}.
Instead, \Crefrange{section-Wirtinger-derivatives}{section-Wirtinger-gradients} introduce Wirtinger derivatives and their relation to real-valued automatic differentiation.
\Cref{section-shortcomings,section-conclusion} conclude this manuscript with a discussion of the shortcomings of the presented approach to complex-valued automatic differentiation and a summary of everything explained in this tutorial.

\paragraph{Target audience.}
We assume a target audience familiar with the basics of analysis and linear algebra and with previous exposure to automatic differentiation, for example, through using \jax{} or \pytorch{}.
The code examples define custom Jacobian-vector and vector-Jacobian products for JAX, and familiarity with this\footnote{\texttt{https://jax.readthedocs.io/en/latest/notebooks/Custom\_derivative\_rules\_for\_Python\_code.html}} guide might be helpful for reading the code examples -- but it is not required.
Unlike related works on complex-valued differentiation, automatic or not, we do not assume previous exposure to complex analysis or differential geometry. 
\Cref{section-related-work} discusses related work more thoroughly.
Setting up complex-valued automatic differentiation without assuming a background in complex analysis was among the main drivers for writing this tutorial.
The only statements from complex analysis that we will rely on are that ``holomorphic'' means ``differentiable'' and that holomorphic functions satisfy the Cauchy--Riemann equations; we refer to, for example, \citet{hunger2007introduction}.
Everything else in this tutorial is linear algebra with complex inner products.

\section{Related work}
\label{section-related-work}

\begin{itemize}
	\item 
	\textit{Background:}
	\Cref{section-background-automatic-differentiation}'s perspective on automatic differentiation follows \jax{}'s \citep{jax2018github} approach to automatic differentiation  via linearisation and transposition \citep{radul2023you}; see also \citet{blondel2024elements}.
	The main reason for choosing this perspective is not that our code examples use \jax{} but rather that this perspective on automatic differentiation uses linear algebra instead of differential geometry.
	We hope a pure linear algebra perspective makes automatic differentiation more accessible to a broader audience than the more mathematically sophisticated perspective using differential geometry.
	\Cref{section-complex-autodiff-as-real-autodiff} essentially spells out what \citet[page 51]{maclaurin2016modeling} writes, which matches how complex differentiation is implemented in \jax{} \citep{jax2018github} and autograd \citep{maclaurin2015autograd} (and likely other environments), but it uses slightly different terminology.

	\item
	\textit{Similar tutorials:}
	\Cref{section-Wirtinger-derivatives} explains Wirtinger derivatives similarly to \citet{kreutz2009complex} (see also \citet{candan2019properly} and \citet{hunger2007introduction}) but we focus on practical advice to implement JVPs and VJPs and skip mathematical (and historical) background wherever possible to dedicate more explanation to JVPs and VJPs.
	Some of our explanations are somewhat new; for example, all of the above texts explain linearisation, the chain rule, and gradients, but none of them discuss VJPs as we do.
	Additionally, our notes mostly avoid topics such as holomorphicity and the Cauchy--Riemann equations while also not assuming that either the in- or outputs of our program are real-valued.
	\citet{lezcano2022automatic} also covers complex automatic differentiation, but not as comprehensively as we do.
	\item	
	\textit{Online discussions:}
	This\footnote{ \texttt{https://discourse.julialang.org/t/taking-complex-autodiff-seriously-in-chainrules/39317/30} } discussion on the Julia forum influenced our presentation of Wirtinger derivatives and their role in Jacobian-vector and vector-Jacobian products. There is also this\footnote{\texttt{https://github.com/tensorflow/tensorflow/issues/3348}} discussion in the TensorFlor issue tracker, this\footnote{ \texttt{https://github.com/HIPS/autograd/blob/master/docs/tutorial.md\#complex-numbers} } explanation of complex numbers in the autograd code, and this\footnote{ \texttt{https://pytorch.org/docs/stable/notes/autograd.html\#autograd-for-complex-numbers} } explanation of complex autograd in the \pytorch{} documentation, all of which motivate a tutorial for complex-valued automatic differentiation like ours: pure JVP and VJP content with minimal complex analysis and minimal differential geometry.
	\item 
	\textit{Research papers:}
	Articles like those by \citet{guo2021scheme,leung1991complex,benvenuto1992complex,nitta1997extension,georgiou1992complex} discuss complex-valued reverse-mode differentiation, but start their discussion with Wirtinger derivatives; we begin with real-valued automatic differentiation and also explain where Wirtinger derivatives come from.
	The matrix cookbook \citep[Equations 234 and 235]{petersen2008matrix} lists similar forward and transposed chain rules to those that we will derive but without much context.
\end{itemize}

\section{Background on automatic differentiation with real numbers}
\label{section-background-automatic-differentiation}

\paragraph{Linearisation and JVPs.}
Automatic differentiation linearises a program by propagating linearisations of functions either forwards or backwards through the computational chain.
Let $V$ and $W$ be finite-dimensional, real-valued vector spaces, for example, $\Rbb$ or $\Rbb^d$.
Complex numbers start in \Cref{section-complex-autodiff-as-real-autodiff}.
We use the notation of $V$ and $W$ instead of simply writing $\Rbb^d$, because we would also like to include spaces like $\Rbb^{4 \times 5}$ or $\Rbb^2 \times \Rbb \times \Rbb^{3 \times 100}$ without ``reshaping'' any inputs.
Let $f: V \rightarrow W$ be a function.
The Jacobian $\partial f(v): V \rightarrow W$ of $f$ is also a function from $V$ to $W$, $\partial f(v): V \rightarrow W$ but linear.
It defines the linearisation of $f$,
\begin{align}
\label{equation-linearisation-of-a-function}
f(v + v') = f(v) + \partial f(v)(v') + {o}(\|v'\|).
\end{align}
Here, ``${o}$'' is the usual ``little O'' notation \citep{burgisser2013condition}.
As users of automatic differentiation, we usually care about Jacobian-vector products (JVPs)
\begin{align}
\label{equation-jacobian-vector-product}
w' = \partial f(v)(v')
\quad
\text{for}
\quad
w = f(v)
\end{align}
and vector-Jacobian products (VJPs; to be defined further below) more than we care about the Jacobian matrix because JVPs and VJPs are cheaper to compute and suffice for most applications.
For example, the linearisation in \Cref{equation-linearisation-of-a-function} requires a single JVP, not a Jacobian matrix, and evaluating the gradient of a scalar-valued function requires a single VJP; more on this later.
Both JVPs and VJPs are essential to automatic differentiation.
We can always reconstruct rows and columns of the Jacobian matrix through JVPs and VJPs with unit basis vectors because, for example, multiplying a matrix with the vector $(0, 1, 0, ..., 0)$ reveals the second column of said matrix.

\paragraph{The chain rule and forward-mode differentiation.}
The JVP satisfies the chain rule in the sense that the JVP of the composition of two functions is a sequence of two JVPs:
Let $f: V \rightarrow W$ and $g: U \rightarrow V$ be differentiable functions, then for all $u, u' \in U$, the chain rule
\begin{align}
\label{equation-chain-rule}
\partial (f \circ g)(u)(u') = \underbrace{ \partial f(v) \underbrace{ \left( \partial g(u) (u') \right)}_{ \text{JVP} }}_{ \text{JVP} }
\end{align}
holds \citep[Equation 3.1]{griewank2008evaluating}.
If a computer program is made up of functions $f$ and $g$, propagating function evaluations and JVPs jointly, according to
\begin{align}
(u, u') \mapsto (v, v') \coloneqq (g(u), \partial g(u)(u')) \mapsto (f(v), \partial f(v)(v')) \mapsto \text{etc.},
\end{align}
instead of propagating only function evaluations $u \mapsto v \mapsto f(v) \mapsto \text{etc.}$ is forward-mode differentiation \citep[Section 3.1]{griewank2008evaluating}.
\Cref{figure-forward-mode-differentiation} includes a code example.
\begin{figure}[t]
\par\bigskip
\begin{center}
\scalebox{0.9}{
\begin{tabular}{ p{0.9\linewidth}}
\begin{lstlisting}[language=Python]
import jax
import jax.numpy as jnp

key = jax.random.PRNGKey(1)
key1, key2 = jax.random.split(key, num=2)

A = jax.random.normal(key1, (3, 3))
x = jax.random.normal(key2, (3,))


def fun(v):
    print("Calling the function, not the JVP...")
    return jnp.dot(v, A @ v)


dfx_ad = jax.jacfwd(fun)(x)  # Calls fun, not JVP


def fun_jvp(primals, tangents):
    print("Calling the JVP, not the function...")
    (v,), (dv,) = primals, tangents
    w = jnp.dot(v, A @ v)
    dw1 = jnp.dot(dv, A @ v)
    dw2 = jnp.dot(v, A @ dv)
    return w, dw1 + dw2


fun = jax.custom_jvp(fun)
fun.defjvp(fun_jvp)
dfx_custom = jax.jacfwd(fun)(x)  # Calls the JVP, not the fun

assert jnp.allclose(dfx_custom, dfx_ad)
\end{lstlisting}
\end{tabular}
}
\caption{Forward-mode differentiation via Jacobian-vector products in \jax{}. 
The inputs to the JVP are called ``tangents'' because technically, $v'$ is in the tangent space of $V$, not $V$ itself \citep{griewank2008evaluating}.
}
\label{figure-forward-mode-differentiation}
\end{center}
\end{figure}

\paragraph{VJPs and reverse-mode differentiation.}
The VJP of $f: V \rightarrow W$ at $v \in V$ is a vector $\overline{v} \in V$, such that for a given $\overline{w} \in W$, and for all $v' \in V$, the identity
\begin{align}
\label{equation-vector-Jacobian-product}
\langle \overline{w}, \partial f(v)(v') \rangle 
= \langle \overline{v}, v' \rangle 
\end{align}
holds \citep{radul2023you}.
We can think of $\overline{w}$ as the ``incoming gradient'' during backpropagation and $\overline{v}$ as the ``outgoing gradient'' that shall be computed.
Later paragraphs will make this interpretation rigorous.
We can also read $\overline{v}$ in \Cref{equation-vector-Jacobian-product} as $\overline{v} = (\partial f(v))^\top\overline{w}$.
However, we define the VJP via \Cref{equation-vector-Jacobian-product} and not via $\partial f(v)^\top \overline{w}$, because \Cref{equation-vector-Jacobian-product} applies to complex values (in principle, but see \Cref{section-complex-autodiff-as-real-autodiff}) and avoids materialising and transposing the Jacobian matrix. 
In practice, using \Cref{equation-vector-Jacobian-product} instead of $\partial f(v)^\top \overline{w}$ more naturally leads to efficient VJP formulas.
The definition of a VJP uses that of a JVP inside an inner product. Thus, where the JVP corresponds to ``linearisation'', the VJP implements ``transposed linearisation''.
The VJP is compositional, just like the JVP, because the VJP of a sequence of operations is a sequence of VJPs:
Let $u, u' \in U$, and $\overline{w} \in W$, be given, abbreviate $v = g(u)$, and let $\overline{v}$ be the same as in \Cref{equation-vector-Jacobian-product}.
Then, the ``transposed chain rule'' holds
\begin{align}
\label{equation-chain-rule-transposed}
\langle \overline{w}, \partial (f \circ g)(u)(u') \rangle 
= \langle \overline{w}, \partial f(g(u)) \partial g(u)(u') \rangle 
= \underbrace{\langle \overline{v} , \partial g(u)(u') \rangle}_{ \text{$1^\text{st}$ VJP } } 
= \underbrace{\langle  \overline{u}  , u' \rangle }_{ \text{$2^\text{nd}$ VJP} }.
\end{align}
Propagating VJPs according to \Cref{equation-chain-rule-transposed} is reverse-mode automatic differentiation \citep{griewank2008evaluating}.
To implement reverse-mode automatic differentiation, we repeatedly apply the transposed chain rule from \Cref{equation-chain-rule-transposed}.
Note how in \Cref{equation-chain-rule-transposed}, the sequence of VJPs reverses the order of function evaluations.
Therefore, unlike forward-mode differentiation, reverse-mode differentiation consists of two passes: One forward pass, which evaluates $u \mapsto v \coloneqq g(u) \mapsto f(g(u))$ and so on, and one backward pass $\overline{w} \mapsto \overline{v}  \mapsto  \overline{u}$ and so on.
The backwards pass depends on all intermediate values of the forward pass, which is why reverse-mode differentiation has a higher memory demand than forward-mode differentiation.

\paragraph{Gradients.}
If the computer program maps $P$ parameters to a scalar, the Jacobian matrix consists of a single row with $P$ columns -- the gradient -- and we can assemble it with a single VJP,
\begin{align}
\langle 1, \partial f(v) (v') \rangle = \langle \overline{v}, v' \rangle
,
\quad
\nabla f (v) \coloneqq \overline{v}
. 
\end{align}
Computing the gradient with a single VJP is why reverse-mode automatic differentiation is so popular for optimising large machine learning models, which map many parameters to a single loss.
\begin{figure}[t]
\par\bigskip
\begin{center}
\scalebox{0.9}{
\begin{tabular}{ p{0.9\linewidth}}
\begin{lstlisting}[language=Python]
import jax
import jax.numpy as jnp

key = jax.random.PRNGKey(1)
key1, key2 = jax.random.split(key, num=2)

A = jax.random.normal(key1, (3, 3))
x = jax.random.normal(key2, (3,))


def fun(v):
    print("Calling the function, not the VJP...")
    return jnp.dot(v, A @ v)


dfx_ad = jax.grad(fun)(x)  # Calls fun, not VJP


def fun_fwd(v):
    print("Calling the forward pass...")
    w = jnp.dot(v, A @ v)
    return w, {"v": v}  # Cache important values


def fun_rev(cache, vjp_incoming):
    print("Calling the backward pass...")
    vjp_outgoing = vjp_incoming * (A + A.T) @ cache["v"]
    return (vjp_outgoing,)


fun = jax.custom_vjp(fun)
fun.defvjp(fun_fwd, fun_rev)
dfx_custom = jax.grad(fun)(x)  # calls VJP, not fun

assert jnp.allclose(dfx_ad, dfx_custom)
\end{lstlisting}
\end{tabular}
}
\caption{Reverse-mode differentiation via vector-Jacobian products in \jax{}. 
}
\label{figure-reverse-mode-differentiation}
\end{center}
\end{figure}
For such mappings with $P$ inputs and one output, we would need $P$ independent JVPs to assemble the gradient because a single JVP reveals only a single column, yet the gradient has one row and $P$ columns.
However, a single VJP reveals a single row of the Jacobian matrix instead of a single column, yielding the full gradient in a single pass.
\Cref{figure-reverse-mode-differentiation} shows a code example.

\paragraph{Challenges with complex numbers.}
Both forward- and reverse-mode differentiation depend on the linearity of the JVP in \Cref{equation-jacobian-vector-product}, which cannot be taken for granted in complex arithmetic unless $f$ is holomorphic \citep{hunger2007introduction} -- this is a critical issue with complex-valued automatic differentiation:
Differentiable programming cannot assume that all functions are holomorphic.
For example, $z \mapsto \Conj{z}$ is not holomorphic, thus even the identity function
\begin{align}
z \mapsto \Conj{z} \mapsto z
\end{align}
is not differentiable if implemented via complex conjugation.
Since the intermediate function $z \mapsto \Conj{z}$ is not differentiable, it does not admit a chain rule, which means that automatic differentiation cannot be defined by applying what we know from real-valued arithmetic.
In other words, even if an ``outer function'' may be holomorphic, unless all ``inner functions'' are, too, automatic differentiation is out of reach.
To solve this problem, we need a notion of differentiability for complex numbers that is less restrictive than holomorphicity but, ideally, reduces to the holomorphic derivative if a function is differentiable.

\section{Complex automatic differentiation via latent real automatic differentiation}
\label{section-complex-autodiff-as-real-autodiff}

\paragraph{Complex numbers are tuples of real numbers.}
Software represents complex numbers as tuples of real numbers.
For example, $z=2 + 3i$ is $(2, 3)$, or $z=0.5-10.1 i$ is $(0.5, -10.1)$.
In light of this representation, one way of defining a meaningful JVP of a complex function is to treat a function of one complex variable as a function of two real variables, differentiate the function, and map the result back into $\Cbb$.
The machine learning literature calls anything that happens in a different, usually higher-dimensional space than the one we are interested in ``latent''.
Therefore, we call this approach to defining a meaningful JVP a ``latent Jacobian-vector product'' because it implements a JVP in latent space (in our case, $\Rbb^2$) and maps the result back into $\Cbb$.
More formally, let $z = x + iy$ and $f(z) = u(x, y) + i v(x, y)$ be known. 
Then, we define the \emph{latent Jacobian-vector product} (latent JVP) as the JVP of $(x, y) \mapsto (u(x, y), v(x,y))$, mapping the result to $\Cbb$, 
\begin{align}
\label{equation-jacobian-vector-product-latent}
\hat{\partial} f(z)(z')
\,=
\begin{pmatrix}
1
&
i
\end{pmatrix}
\begin{pmatrix}
\partial_x u & \partial_y u
\\
\partial_x v & \partial_y v
\end{pmatrix}
\begin{pmatrix}  
x'
\\
y'
\end{pmatrix}
.
\end{align}
If the partial derivatives of $u$ and $v$ are well-defined, the latent JVP in \Cref{equation-jacobian-vector-product-latent} is well-defined, which is far less restrictive than assuming that $f$ is holomorphic.
Propagating latent JVPs according to \Cref{equation-jacobian-vector-product-latent} is forward-mode differentiation for complex numbers. (Reverse mode waits until \Cref{section-Wirtinger-gradients}.)
The latent JVP is not linear in $z'$, only in $x'$ and $y'$ -- that is, \emph{after} separating real and imaginary components.
It is important that we call \Cref{equation-jacobian-vector-product-latent} a ``latent JVP'', not a ``JVP'', because the latent JVP only reduces to the JVP if $f$ is holomorphic; more on this later.
For the same reason, we use the notation $\hat\partial$ instead of $\partial$.
Regardless, the latent JVP constructs automatic differentiation consistent with real-valued derivatives and holomorphic functions as follows.

\paragraph{Consistency with real-valued derivatives.}
By construction, $\hat\partial f$ is consistent with real-valued differentiation in the sense that we can take an existing real-valued automatic differentiation environment, introduce a complex number type as a tuple of the real and imaginary parts of a complex number, and define the JVP on this type using real-valued JVPs (\Cref{equation-jacobian-vector-product}).
In this scenario, mapping results from $\Rbb^2$ to $\Cbb$ essentially reduces to a string representation of these specific $\Rbb^2$ tuples because all arithmetic happens in $\Rbb^2$. 
Consistency with real-valued differentiation is helpful because it enables a user to switch between real- and complex-valued differentiation without having to redefine the concept of a directional derivative, which saves a considerable amount of work (but at a price; see the discussion in \Cref{section-shortcomings}).

\paragraph{Consistency with holomorphicity.}
The latent JVP in \Cref{equation-jacobian-vector-product-latent} is not a linear function of $z'$, only of $x'$ and $y'$ as separate inputs, which is why we don't call it a JVP.
However, if $f$ is holomorphic, the latent JVP reduces to the JVP from \Cref{equation-jacobian-vector-product} because if $f$ is holomorphic, $u$ and $v$ satisfy the Cauchy--Riemann equations \citep[e.g.][]{hunger2007introduction}, 
\begin{align}
\partial_x u = \partial_y v , \quad \partial_y u = - \partial_x v,
\end{align}
in which case the JVP simplifies:
\begin{subequations}
\begin{align}
\hat\partial f(z)(z')
&=
\begin{pmatrix}
1
&
i
\end{pmatrix}
\begin{pmatrix}
\partial_x u & \partial_y u
\\
\partial_x v & \partial_y v
\end{pmatrix}
\begin{pmatrix}
x'
\\
y'
\end{pmatrix}
\\
&=
\begin{pmatrix}
1
&
i
\end{pmatrix}
\begin{pmatrix}
\partial_x u & \partial_y u
\\
-\partial_y u & \partial_x u
\end{pmatrix}
\begin{pmatrix}
x'
\\
y'
\end{pmatrix}
\\
&=
\begin{pmatrix}
\partial_x u -i \partial_y u
&
\partial_y u + i \partial_x u
\end{pmatrix}
\begin{pmatrix}
x'
\\
y'
\end{pmatrix}
\\
&=
(\partial_x u - i \partial_y u)
(x'+  i y')
.
\end{align}
\end{subequations}
In other words, if $f$ is holomorphic, the latent JVP matches the JVP because it multiplies an input $z'$ with the Jacobian $\partial f(z) = \partial_x u - i \partial_y u$.
However, and this is crucial because we cannot assume holomorphicity of all functions that make up a complex-valued computer program, the latent JVP in \Cref{equation-jacobian-vector-product-latent} is well-defined even if $f$ is not holomorphic.

\paragraph{Two examples.}
Here are two examples of latent JVPs: one for a holomorphic function and one for a non-holomorphic function.
First, the function $f(z) = 2z$ is holomorphic and separates into $u(x,y) = 2x$ and $v(x,y) = 2y$.
Its latent JVP is thus
\begin{align}
\hat\partial f(z)(z') =
\begin{pmatrix}
1 & i
\end{pmatrix}
\begin{pmatrix}
2 & 0 \\
0 & 2
\end{pmatrix}
\begin{pmatrix}
x'
\\
y'
\end{pmatrix}
=
2
\begin{pmatrix}
1 & i
\end{pmatrix}
\begin{pmatrix}
x'
\\
y'
\end{pmatrix}
= 2 z'.
\end{align}
The latent JVP is a linear function of $z'$, which means that it is the (actual) JVP.
The Jacobian of $f$ equals $2$, which matches our expectation because $f$ scales inputs by factor $2$.
For the non-holomorphic example, let $\Conj{z}$ be the complex conjugate of $z$ and let $A^\top$ be the transpose of $A$.
This tutorial always spells out conjugate transposes as $\Conj{A}^\top$.
For $f(z) = \Conj{z}^\top A z$, we separate the real and imaginary parts as
\begin{align}
f(z) = (x - iy)^\top A  (x+iy)
= (x^\top A x + y^\top A y) + i (x^\top A y - y^\top A x) 
\end{align}
with $u(x,y) = x^\top A x + y^\top A y$ and $v(x,y) = x^\top Ay - y^\top A x$,
which implies the partial derivatives
\begin{align}
\partial_x u = x^\top(A + A^\top), 
\quad
\partial_y u = y^\top (A + A^\top),
\quad
\partial_x v = y^\top (A^\top - A),
\quad
\partial_y v = x^\top (A - A^\top).
\end{align}
Those partial derivatives, via \Cref{equation-jacobian-vector-product-latent}, lead to the latent JVP
\begin{subequations}
\label{equation-example-jvp-of-inner-product}
\begin{align}
\hat\partial f(z)(z') 
&= 
\begin{pmatrix} 1 & i \end{pmatrix}
\begin{pmatrix}
x^\top (A + A^\top)  & y^\top (A + A^\top)
\\
y^\top (A^\top - A ) & x^\top (A - A^\top)
\end{pmatrix}
\begin{pmatrix}
x'
\\
y'
\end{pmatrix}
\\
&=
\left[ x^\top (A + A^\top) x' + y^\top (A + A^\top) y'\right]
+ i\left[y^\top (A^\top-A) x' + x^\top (A - A^\top) y'
\right]
\\
&=
(x-iy)^\top A (x'+iy')
+
(x+iy)^\top A^\top (x'-iy')
\\
&= 
\Conj{z}^\top A z' + z^\top A^\top \Conj{z'}.
\end{align}
\end{subequations}
The latent JVP matches what we might have expected from applying an ``intuitive chain rule'' to $f(z) = \Conj{z}^\top A z$, but \Cref{equation-jacobian-vector-product-latent} (via \Cref{equation-example-jvp-of-inner-product}) makes this intuitive chain rule rigorous.
Note how the latent JVP in \Cref{equation-example-jvp-of-inner-product} is not linear in $z'$, because it involves $\Conj{z'}$.
The function $f(z) = \Conj{z}^\top A z$ is not holomorphic.
\Cref{figure-forward-mode-differentiation-complex} turns \Cref{equation-example-jvp-of-inner-product} into \jax{} code.
\begin{figure}[t]
\par\bigskip
\begin{center}
\scalebox{0.9}{
\begin{tabular}{ p{0.9\linewidth}}
\begin{lstlisting}[language=Python]
import jax
import jax.numpy as jnp

key = jax.random.PRNGKey(1)
key1, key2, key3 = jax.random.split(key, num=3)

A = jax.random.normal(key1, (3, 3), dtype=complex)
x = jax.random.normal(key2, (3,), dtype=complex)
dx = jax.random.normal(key3, (3,), dtype=complex)


def fun(v):
    print("Calling the function, not the JVP...")
    return jnp.dot(v.conj(), A @ v)


_, dfx_ad = jax.jvp(fun, (x,), (dx,))  # Calls fun, not JVP


def fun_jvp(primals, tangents):
    print("Calling the JVP, not the function...")
    (v,), (dv,) = primals, tangents
    w = jnp.dot(v.conj(), A @ v)
    dw1 = jnp.dot(dv.conj(), A @ v)
    dw2 = jnp.dot(v.conj(), A @ dv)
    return w, dw1 + dw2


fun = jax.custom_jvp(fun)
fun.defjvp(fun_jvp)
_, dfx_custom = jax.jvp(fun, (x,), (dx,))  # Calls JVP, not fun

assert jnp.allclose(dfx_ad, dfx_custom)
\end{lstlisting}
\end{tabular}
}
\caption{Complex forward-mode differentiation in \jax{}.
Unlike the script in \Cref{figure-forward-mode-differentiation}, this script works with complex numbers.
}
\label{figure-forward-mode-differentiation-complex}
\end{center}
\end{figure}

\paragraph{So what's the problem?}
The latent JVP appears to be a well-defined extension of the JVP to complex values.
However, the expression of the latent JVP in \Cref{equation-jacobian-vector-product-latent} isn't enough: for a given function $f$, identifying $u$ and $v$ and its partial derivatives may require more work than we are willing to do. For example, determining the latent JVP of $f(z) = \Conj{z}^\top A z$ as above, via separating $f$'s real and imaginary components, required about half a page of arithmetic, whereas its real-valued equivalent needs only one or two lines.
Now, try computing the latent JVP of a function like $f(z) = z^5 \Conj{z}^{4}$ with the same technique -- this will be a lot of work.
Fortunately, a slight reformulation of how we express latent JVPs simplifies their derivation considerably.
This simplification will be the subject of the rest of this tutorial.

\section{Wirtinger derivatives}
\label{section-Wirtinger-derivatives}

\paragraph{The need for a new basis.}
It turns out the above derivation was only laborious because our coordinate system was, loosely speaking, ``bad at chain rules'': it is tedious to identify how, for example, the imaginary component of $f$ depends on the real-valued component $z$.
To see this, we will change the basis of $\Rbb^2$ and obtain a much simpler expression for latent JVPs than \Cref{equation-jacobian-vector-product-latent}.

\paragraph{Basis change.}
\Cref{section-complex-autodiff-as-real-autodiff} uses that a complex number is a tuple of two real numbers to derive the latent JVP in \Cref{equation-jacobian-vector-product-latent}.
For example, \Cref{section-complex-autodiff-as-real-autodiff} represents the value $z = 2 + 3i$ as $(2,3)$. 
More technically, representing $\Cbb$ with two real numbers regards $\Cbb$ as a vector space over $\Rbb$ (as opposed to a vector space over $\Cbb$), which makes $\Cbb$ two-dimensional with the basis $\{1, i\}$.
However, even though $\{1, i\}$ is the standard basis, it is not the only one: 
Instead of $\{1, i\}$, we could also consider $\{x + yi, x - yi\} = \{z, \Conj{z}\}$ as a basis (using $z$ as a basis element for representing $z$ might seem like a circular argument, but it will make sense in a minute).
Then, we can represent
\begin{align}
\label{equation-representation-in-wirtinger-basis}
z
&=
\begin{pmatrix}
1 & i
\end{pmatrix}
\begin{pmatrix}
x
\\
y
\end{pmatrix}
=
\frac{1}{2}
\begin{pmatrix}
1 & i
\end{pmatrix}
\begin{pmatrix}
1 & 1
\\
-i & i
\end{pmatrix}
\begin{pmatrix}
1 & i
\\
1 & -i
\end{pmatrix}
\begin{pmatrix}
x
\\
y
\end{pmatrix}
=
\begin{pmatrix}
1 & 0
\end{pmatrix}
\begin{pmatrix}
x + iy
\\
x-iy
\end{pmatrix}
.
\end{align}
\Cref{equation-representation-in-wirtinger-basis} is also an expansion for $z$ with real coefficients (zero and one) and complex basis elements ($x+iy$ and $x-iy$).
This basis expansion is somewhat redundant because one of the coefficients is zero, but it is a valid basis for $\Cbb$ as a vector space over $\Rbb$.
Inside a latent JVP, the coefficients will no longer cancel each other, and the basis representation will become more interesting.

\paragraph{Wirtinger derivatives.}
Using the basis change from \Cref{equation-representation-in-wirtinger-basis}, the latent JVP from \Cref{equation-jacobian-vector-product-latent} gains a new representation,
\begin{subequations}
\label{equation-wirtinger-derivation}
\begin{align}
\hat\partial f(z)(z')
&=
\begin{pmatrix}
1
&
i
\end{pmatrix}
\begin{pmatrix}
\partial_x u & \partial_y u
\\
\partial_x v & \partial_y v
\end{pmatrix}
\begin{pmatrix}
x'
\\
y'
\end{pmatrix}
\\
&=
\frac{1}{2}
\begin{pmatrix}
1
&
i
\end{pmatrix}
\begin{pmatrix}
\partial_x u & \partial_y u
\\
\partial_x v & \partial_y v
\end{pmatrix}
\begin{pmatrix}
1 & 1
\\
-i & i
\end{pmatrix}
\begin{pmatrix}
1 & i
\\
1 & -i
\end{pmatrix}
\begin{pmatrix}
x'
\\
y'
\end{pmatrix}
\\
&=
\frac{1}{2}
\begin{pmatrix}
\partial_x u - i \partial_y u + i(\partial_x v - i \partial_y v)
&
\partial_x u + i \partial_y u + i(\partial_x v + i \partial_y v)
\end{pmatrix}
\begin{pmatrix}
x' +i y'
\\
x' - iy'
\end{pmatrix}
\\
&=
\begin{pmatrix}
\partial_z f
&
\partial_\Conj{z} f
\end{pmatrix}
\begin{pmatrix}
z'
\\
\Conj{z'}
\end{pmatrix}
\end{align}
\end{subequations}
which introduces the new derivatives
\begin{subequations}
\label{equation-wirtinger-derivatives}
\begin{align}
\partial_z f 
&\coloneqq 
\frac{1}{2} (\partial_x u - i \partial_y u + i(\partial_x v - i \partial_y v))
=
\frac{1}{2}
( \partial_x f - i \partial_y f)
\\
\partial_\Conj{z} f 
&\coloneqq
\frac{1}{2}(
\partial_x u + i \partial_y u + i(\partial_x v + i \partial_y v))
=
\frac{1}{2}
( \partial_x f + i \partial_y f).
\end{align}
\end{subequations}
These new derivatives' subscripts are chosen to match their direction, which means that $\partial_z f$ gets its label for corresponding to $z' = x' + iy'$, and $\partial_\Conj{z}f$ gets its label for corresponding to $\Conj{z'} = x' - iy'$.
The derivatives in \Cref{equation-wirtinger-derivatives} are called ``Wirtinger derivatives'' \citep{wirtinger1927formalen}.
According to \Cref{equation-wirtinger-derivation}, Wirtinger derivatives allow expressing the latent JVP as simple as
\begin{align}
\label{equation-jacobian-vector-product-latent-wirtinger}
\hat\partial f(z)(z') = \partial_z f(z)(z') + \partial_\Conj{z} f (z)(\Conj{z'}).
\end{align}
\Cref{equation-jacobian-vector-product-latent-wirtinger} is equivalent to \Cref{equation-jacobian-vector-product-latent}. Thus, propagating latent JVPs forward through the computational chain according to either \Cref{equation-jacobian-vector-product-latent} or \Cref{equation-jacobian-vector-product-latent-wirtinger} (whichever is more convenient) is forward-mode automatic differentiation with complex numbers.
The difference between \Cref{equation-jacobian-vector-product-latent,equation-jacobian-vector-product-latent-wirtinger} is a different basis of $\Rbb^2$ ``under the hood'', but both expressions work with the same in- and outputs.
Like the latent JVP itself, Wirtinger derivatives are well-defined whenever a function's real and imaginary parts are real-differentiable, which is strictly less restrictive than holomorphicity. For example, $f(z) = \Conj{z}^\top A z$ admits Wirtinger derivatives, even though it is not holomorphic.

\paragraph{Working with Wirtinger derivatives.}
Whenever we work with Wirtinger derivatives, $z$ and $\Conj{z}$ are treated as independent variables.
That means that to compute $\partial_z f$, we fix $\Conj{z}$ and differentiate with respect to $z$, and vice versa.
While from a technical point of view, this might be somewhat confusing because $z$ and $\Conj{z}$ are not independent variables, the following examples show how naturally this separation occurs in practice:
We return to latent JVPs of $f(z) = \Conj{z}^\top A z$. 
The Wirtinger derivatives of $f$ are
\begin{align}
\label{equation-wirtinger-derivatives-inner-product}
\partial_z f = \Conj{z}^\top A, 
\quad 
\partial_\Conj{z} f = z^\top A^\top,
\end{align}
because we treat $z$ and $\Conj{z}$ as independent variables.
\Cref{equation-wirtinger-derivatives-inner-product} implies the latent JVP
\begin{align}
\label{equation-latent-jvp-inner-product-wirtinger}
\hat\partial f(z)(z') = \Conj{z}^\top A z' + \Conj{z'}^\top \Conj{A}^\top z.
\end{align}
This latent JVP matches the latent JVP in \Cref{equation-example-jvp-of-inner-product} in \Cref{section-complex-autodiff-as-real-autodiff}, but was considerably easier to derive.
As another example, consider $f(z) = z^5\Conj{z}^4$, which has Wirtinger derivatives 
\begin{align}
\partial_z f(z) = 5 z^4 \Conj{z}^4,
\quad
\partial_\Conj{z} f(z) = 4 z^5 \Conj{z}^3,
\end{align}
which in turn leads to the latent JVP 
\begin{align}
\label{equation-latent-jvp-complicated-function-wirtinger}
\hat\partial f(z)(z') = 5 z^4 \Conj{z}^4 z' + 4 z^5 \Conj{z}^3 \Conj{z'}.
\end{align}
Deriving the same rule by identifying real and complex parts of $f$ and their partial derivatives would have been significantly more laborious.
The rules in \Cref{equation-latent-jvp-complicated-function-wirtinger,equation-latent-jvp-inner-product-wirtinger} match what we might have done intuitively; Wirtinger calculus makes this intuition rigorous.

\paragraph{Wirtinger derivatives in software.}
In software, there is no difference between implementing latent JVPs through \Cref{equation-jacobian-vector-product-latent-wirtinger} or \Cref{equation-jacobian-vector-product-latent} because \Cref{equation-jacobian-vector-product-latent-wirtinger,equation-jacobian-vector-product-latent} are simply two different ways of deriving the required expression.
For example, \Cref{figure-forward-mode-differentiation-complex} remains correct independent of whether we get the expression from \Cref{equation-jacobian-vector-product-latent-wirtinger} or \Cref{equation-jacobian-vector-product-latent}.
Both ways, Wirtinger or not, enable forward-mode automatic differentiation with complex numbers.
For example, the operation $z \mapsto \Conj{z} \mapsto z$ from the end of \Cref{section-background-automatic-differentiation} was not compatible with the chain rule because $z \mapsto \Conj{z}$ is not differentiable.
Nonetheless, the latent JVP of $z \mapsto \Conj{z} \mapsto z$ is
$z' \mapsto \Conj{z'} \mapsto z'$, which implements the identity function as expected.
Since the ``outer function'' of $z \mapsto \Conj{z} \mapsto z$ is holomorphic, the latent JVP reduces to the true JVP -- even though none of the ``inner functions'' were holomorphic.
So, in summary, the latent JVP enables a meaningful formulation of forward-mode differentiation with complex numbers because it provides a chain rule  (\Cref{equation-jacobian-vector-product-latent,equation-jacobian-vector-product-latent-wirtinger}) that does not require holomorphicity.
Choosing between the two formulations of latent JVPs is a matter of personal preference; usually, the route via Wirtinger derivatives is less work.

\section{Vector-Jacobian products}
\label{section-vector-Jacobian-products}

\paragraph{From latent JVPs to latent VJPs.}
Now that we know how to implement latent JVPs, what about VJPs?
Like in \Cref{section-complex-autodiff-as-real-autodiff}, ``actual'' VJPs are not an option because they would require holomorphicity of the function $f$.
We turn to a latent VJP.
However, constructing a latent VJP is more difficult than applying transposition like in \Cref{equation-vector-Jacobian-product} to the latent JVP $\hat\partial f$.
The reason for this difficulty is that if $f$ is not holomorphic, $\hat\partial f(z)(z')$ is not linear in $z'$, which means that the operator $\hat\partial f(z)$ does not have a well-defined adjoint.
That said, we can define a latent VJP by computing the VJP of $f$ as a function from $\Rbb^2$ to $\Rbb^2$ and map the result into $\Cbb$.

\paragraph{Latent VJP.}
Recall $z = x + iy$ and $y \coloneqq f(z) = u(x,y) + i v(x,y)$ from \Cref{section-complex-autodiff-as-real-autodiff}.
Let $\overline{f} = \overline{u} \pm i \,\overline{v}$ be a vector with the same shape as the output of $f$ (we don't commit to a sign in front of $i \overline{v}$ at this point; see the next paragraph).
We define a latent VJP as the result of the following two-step procedure:
first, finding $\overline{x}$ and $\overline{y}$ that satisfy the transposition
\begin{align}
\label{equation-wirtinger-linearisation-transposed-preliminary}
\left\langle
\begin{pmatrix}
\overline{u}
\\
\overline{v}
\end{pmatrix}
,
\begin{pmatrix}
\partial_x u & \partial_y u
\\
\partial_x v & \partial_y v
\end{pmatrix}
\begin{pmatrix}
x'
\\
y'
\end{pmatrix}
\right\rangle
=
\left\langle
\begin{pmatrix}
\overline{x}
\\
\overline{y}
\end{pmatrix}
,
\begin{pmatrix}
x'
\\
y'
\end{pmatrix}
\right\rangle
\end{align}
and second, setting $\overline{z} = \overline{x} \pm i \,\overline{y}$.
We call a latent VJP against $\overline{f} = 1$ a latent gradient.
Both $\overline{f}$ and $\overline{z}$ have an undecided sign in front of their imaginary parts.
This sign is undecided because, unlike for the latent JVP, choosing the mapping from $\Rbb^2$ to $\Cbb$ does not have an intuitively correct answer for the latent VJP.
Loosely speaking, since we map the ``left'' factor of the inner product to and back from $\Rbb^2$ instead of the ``right'' factor, we might prefer the complex conjugate of the mapping from \Cref{section-Wirtinger-derivatives}: $(1, -i)$ instead of $(1, i)$. 
The following discussion shows how both are reasonable solutions.

\paragraph{Conjugate or not?}
Denote $\overline{z} = \overline{x} + i c_1 \overline{y}$ and $\overline{f} = \overline{u} + i c_2 \overline{v}$, with $c_1, c_2 \in \{1, -1\}$.
We compute the latent VJP of two holomorphic functions to gain an intuition for the choices of $c_1$ and $c_2$.
We choose holomorphic functions because, for holomorphic functions, linearisation and transposition are well-defined, so we know how we would like the latent VJP to behave from combining the real-valued VJP (\Cref{section-background-automatic-differentiation}) with a complex inner product.
For example, the VJP of the identity $f(z) = z$ should be the identity because its Jacobian is equal to one and thus self-adjoint. 
For $f(z) = z$, the definition of the latent VJP implies, $\overline{x} = \overline{u}$ and $\overline{y} = c_1 \overline{v}$, which leads to
\begin{align}
\overline{z} = \overline{u} + i \,c_1 c_2 \overline{v}.
\end{align}
To make this latent VJP behave like the identity function, $c_1$ and $c_2$ must be equal, and we say $c_1 = c_2 = c$ with $c \in \{1, -1\}$.
To choose $c$, we take the latent VJP of the holomorphic $f(z) = \frac{1}{2} z^2$,
\begin{align}
\label{equation-latent-vjp-of-square}
\left\langle
\begin{pmatrix}
\overline{u}
\\
c \overline{v}
\end{pmatrix}
,
\begin{pmatrix}
x & -y \\
y & x
\end{pmatrix}
\begin{pmatrix}
x'
\\
y'
\end{pmatrix}
\right\rangle
=
\left\langle
\begin{pmatrix}
x & y \\
-y & x
\end{pmatrix}
\begin{pmatrix}
\overline{u}
\\
c \overline{v}
\end{pmatrix}
,
\begin{pmatrix}
x'
\\
y'
\end{pmatrix}
\right\rangle
=
\left\langle
\begin{pmatrix}
x \overline{u} + c y \overline{v}
\\
 -y \overline{u} + c x \overline{v}
\end{pmatrix}
,
\begin{pmatrix}
x'
\\
y'
\end{pmatrix}
\right\rangle
.
\end{align}
\Cref{equation-latent-vjp-of-square} identifies $\overline{x}$ and $\overline{y}$, thus the latent VJP is 
\begin{align}
\overline{z} 
= 
\overline{x} + i \, c \, \overline{v}
=
(x - i c y)(\overline{u} + i \,\overline{v}).
\end{align}
Depending on whether we choose $c=1$ or $c=-1$, the latent VJP is either $\overline{z} = \Conj{z} \overline{f}$ (choose $c=1$) or $\overline{z} = z \overline{f}$ (choose $c=-1$).
Thus, the latent gradient is either $\nabla f(z) = \Conj{z}$ or $\nabla f(z) = z$.
Either option is fine, and there are good reasons for both:
\begin{itemize}
	\item Choosing $\overline{z} = z \overline{f}$, which corresponds to $c=-1$, emphasises that the latent VJP should implement an actual vector-Jacobian product, as opposed to a vector-conjugate-Jacobian product because the Jacobian of $f$ is $z$, not $\Conj{z}$.
	This naming consistency is important when writing software because when users call a function called ``vjp'', they expect a vector-Jacobian product -- independent of real or complex arithmetic. 
	\item Choosing $\overline{z} = \Conj{z} \overline{f}$, which corresponds to $c=1$, emphasises that the VJP should implement the adjoint of the JVP, which matches the definition of the real-valued VJP as the adjoint of the real-valued JVP (\Cref{equation-vector-Jacobian-product}).
	It is not a literal vector-Jacobian product, but it leads to the gradients we would expect from computing adjoints of columns of the Jacobian matrix: $\nabla f(z) = \Conj{z}$ because $\partial f(z) = z$.

\end{itemize}
In conclusion, when defining reverse-mode automatic differentiation via latent VJPs, we must sacrifice exactly one of the following: 
(i) latent VJPs implement vector-Jacobian products (as opposed to vector-conjugate-Jacobian products), 
(ii) latent gradients reduce to the correct gradients for holomorphic functions, 
(iii) latent gradients are latent VJPs against a unit vector.
In real arithmetic, this kind of friction doesn't exist because the adjoint of a linear operator does not involve complex conjugation.
This phenomenon underlines how complex values might not have been a critical consideration when choosing the terminology of automatic differentiation.
We adopt the convention of $c=1$, which leads to the latent VJP $\overline{z} = \Conj{z} \overline{f}$ of $f(z) = \frac{1}{2} z^2$ and the latent gradient $\nabla f(z) = \Conj{z}$.
This choice prioritises a correct gradient and accepts that the name ``latent VJP'' becomes inaccurate since it technically does not implement a vector-Jacobian product.
Regardless, we don't change the name of the latent VJP because we are used to ``VJP'' meaning ``reverse-mode'' from real-valued automatic differentiation.

\paragraph{Conjugate or not: software.}
As for software, it is sometimes a bit unclear which of the two conventions a given environment implements.
This\footnote{ \texttt{https://github.com/tensorflow/tensorflow/issues/3348} } discussion on \tensorflow{}'s, this\footnote{ \texttt{https://github.com/pytorch/pytorch/issues/41857} } one on \pytorch{}'s, and this\footnote{ \texttt{https://github.com/google/jax/issues/4891} } one on \jax{}'s issue board are examples.
Fortunately, it can be relatively straightforward to check a software environment's choice by evaluating the gradient of $f(z) = \frac{1}{2} z^2$ at $z=1+i$: if the result is $1-i$, the ``plus'' convention holds but if the result is $1+i$, the ``minus'' convention holds.
See the script in \Cref{figure-code-display-gradient-conventions}, whose results are in \Cref{table-conjugate-or-not-conventions-software}.
Notably, at the time of writing this tutorial, \jax{} adopts a different convention from \pytorch{} and \tensorflow{}; see this\footnote{ \texttt{https://github.com/google/jax/issues/4891}
 } discussion on \jax{}'s issue board.
\begin{figure}[t]
\par\bigskip
\begin{center}
\scalebox{0.9}{
\begin{tabular}{ p{0.9\linewidth}}
\begin{lstlisting}[language=Python]
import jax
import tensorflow as tf
import torch


def fun(z):
    """Evaluate the test-function."""
    return z**2 / 2.0


x0 = 1.0 + 1j  # the test-location

print(f"JAX:\n\t{jax.__version__}", end="\n\t")
print(jax.grad(fun, holomorphic=True)(x0))


print(f"PyTorch:\n\t{torch.__version__}", end="\n\t")
x = torch.tensor(x0, requires_grad=True)
y = fun(x)
y.backward(gradient=torch.tensor(1.0 + 0 * 1j))
print(x.grad)

print(f"TensorFlow:\n\t{tf.__version__}", end="\n\t")
x = tf.Variable(x0)
with tf.GradientTape() as tape:
    y = fun(x)
print(tape.gradient(y, x).numpy())
\end{lstlisting}
\end{tabular}
}
\caption{
	Evaluate the gradient of $f(z) = \frac{1}{2} z^2$ to reveal the gradient convention.
}
\label{figure-code-display-gradient-conventions}
\end{center}
\end{figure}
\begin{table}[ht]
\caption{Gradient of $f(z) = \frac{1}{2}z^2$ evaluated at $z=1+i$.}
\label{table-conjugate-or-not-conventions-software}
\begin{center}
\begin{tabular}{ p{0.15\linewidth} p{0.15\linewidth} p{0.15\linewidth} p{0.15\linewidth}}
\toprule
& \jax{} & \pytorch{} & \tensorflow{} 
\\
\midrule
Version:
	& \tt 0.4.31
	& \tt 2.4.1+cpu
	& \tt 2.17.0
\\
$\nabla f(1+i)$: 
	& 
	$1 + i$
	& 
	$1 - i$
	& 
	$1 - i$
\\
\bottomrule
\end{tabular}
\end{center}
\end{table}
The derivations below follow \tensorflow{} and \pytorch{} by using the ``plus'' convention, which means $\overline{f} = \overline{u} + i \, \overline{v}$ and $\overline{z} = \overline{x} + i \overline{y}$, even though our examples use \jax{}.
The scripts at the end of this section will demonstrate the ease of switching conventions when implementing latent VJPs.

\section{Latent VJPs with Wirtinger derivatives}
\label{section-Wirtinger-gradients}

\paragraph{Transposing Wirtinger derivatives.}
Even though the latent VJP in \Cref{equation-wirtinger-linearisation-transposed-preliminary} is well-defined and consistent with real-valued differentiation and holomorphic functions like the latent JVP (the same arguments apply), it is not optimal for functions whose $u$ and $v$ are laborious to identify.
Like in \Cref{section-Wirtinger-derivatives}, the solution is to replace partial derivatives like $\partial_x u$ with Wirtinger derivatives.
To describe latent VJP in terms of Wirtinger derivatives, rewrite \Cref{equation-wirtinger-linearisation-transposed-preliminary} using the same basis change as in \Cref{section-Wirtinger-derivatives} to both sides of the inner product,
\begin{subequations}
\label{equation-wirtinger-transposed-linearisation-basis-change}
\begin{align}
&
\left\langle
\begin{pmatrix}
\overline{u}
\\
\overline{v}
\end{pmatrix}
,
\begin{pmatrix}
\partial_x u & \partial_y u
\\
\partial_x v & \partial_y v
\end{pmatrix}
\begin{pmatrix}
x'
\\
y'
\end{pmatrix}
\right\rangle
\\
&
=
\frac{1}{2}
\left\langle
\begin{pmatrix}
1 & 1
\\
-i & i
\end{pmatrix}
\begin{pmatrix}
1 & i
\\
1 & -i
\end{pmatrix}
\begin{pmatrix}
\overline{u}
\\
\overline{v}
\end{pmatrix}
,
\frac{1}{2}
\begin{pmatrix}
\partial_x u & \partial_y u
\\
\partial_x v & \partial_y v
\end{pmatrix}
\begin{pmatrix}
1 & 1
\\
-i & i
\end{pmatrix}
\begin{pmatrix}
1 & i
\\
1 & -i
\end{pmatrix}
\begin{pmatrix}
x'
\\
y'
\end{pmatrix}
\right\rangle
\\
&
=
\frac{1}{2}
\left\langle
\begin{pmatrix}
\overline{f}
\\
\Conj{\overline{f}}
\end{pmatrix}
,
\frac{1}{2}
\begin{pmatrix}
1 & i
\\
1 & -i
\end{pmatrix}
\begin{pmatrix}
\partial_x u & \partial_y u
\\
\partial_x v & \partial_y v
\end{pmatrix}
\begin{pmatrix}
1 & 1
\\
-i & i
\end{pmatrix}
\begin{pmatrix}
z'
\\
\Conj{z'}
\end{pmatrix}
\right\rangle
\\
&
=
\frac{1}{2}
\left\langle
\begin{pmatrix}
\overline{f}
\\
\Conj{\overline{f}}
\end{pmatrix}
,
\begin{pmatrix}
\partial_z f & \partial_\Conj{z} f
\\
\Conj{\partial_\Conj{z} f} & \Conj{\partial_{z} {f}}
\end{pmatrix}
\begin{pmatrix}
z'
\\
\Conj{z'}
\end{pmatrix}
\right\rangle
.
\end{align}
\end{subequations}
To change the gradient convention, exchange $\overline{f}$ and $\Conj{\overline{f}}$.
The central difference between \Cref{equation-wirtinger-transposed-linearisation-basis-change} and \Cref{equation-jacobian-vector-product-latent-wirtinger} is that \Cref{equation-wirtinger-transposed-linearisation-basis-change} involves the Wirtinger derivatives of $f$ and $\Conj{f}$, and \Cref{equation-jacobian-vector-product-latent-wirtinger} only of $f$.
(The Wirtinger derivatives of $\Conj{f}$ and $f$ relate through complex conjugation, which is used in the last step of \Cref{equation-wirtinger-transposed-linearisation-basis-change}.)
Reorder the terms in \Cref{equation-wirtinger-transposed-linearisation-basis-change},
\begin{subequations}
\label{equation-deriving-xi}
\begin{align}
&
\frac{1}{2}
\left\langle
\begin{pmatrix}
\overline{f}
\\
\Conj{\overline{f}}
\end{pmatrix}
,
\begin{pmatrix}
\partial_z f & \partial_\Conj{z} f
\\
\Conj{\partial_\Conj{z} f} & \Conj{\partial_{z} {f}}
\end{pmatrix}
\begin{pmatrix}
z'
\\
\Conj{z'}
\end{pmatrix}
\right\rangle
\\
&\quad\quad
=
\frac{1}{2}
\left\langle
\begin{pmatrix}
\Conj{\partial_z f}^\top
&
(\partial_\Conj{z} f)^\top
\\
\Conj{\partial_\Conj{z} f}^\top
&
(\partial_z f)^\top
\end{pmatrix}
\begin{pmatrix}
\overline{f}
\\
\Conj{\overline{f}}
\end{pmatrix}
,
\begin{pmatrix}
z'
\\
\Conj{z'}
\end{pmatrix}
\right\rangle
\\
&\quad\quad
=
\frac{1}{2}
\left\langle
\begin{pmatrix}
\xi
\\
\Conj{ \xi }
\end{pmatrix}
,
\begin{pmatrix}
z'
\\
\Conj{z'}
\end{pmatrix}
\right\rangle
\end{align}
\end{subequations}
with a new variable $\xi$, defined as
\begin{align}
\label{equation-definition-xi}
\xi \coloneqq 
\Conj{\partial_{z} f}^\top \overline{f}+ (\partial_\Conj{z} f)^\top  \Conj{ \overline{f} }
.
\end{align}
According to the symmetry in \Cref{equation-deriving-xi}, either $\xi$ or $\Conj{\xi}$ uniquely represent the latent VJP.
The correct choice between $\xi$ and $\Conj{\xi}$ depends on the gradient convention.

\paragraph{Reverse-mode differentiation.}
Recall that our original plan for constructing latent VJPs was to identify $\overline{x}$ and $\overline{y}$ according to \Cref{equation-wirtinger-linearisation-transposed-preliminary} and to map the result to the complex numbers via $\overline{z} = \overline{x} + i\, \overline{y}$.
Now that we know $\xi$ and $\Conj{\xi}$, we can identify $\overline{x}$ and $\overline{y}$:
\begin{subequations}
\label{equation-identifying-xbar-and-ybar}
\begin{align}
\frac{1}{2}
\left\langle
\begin{pmatrix}
\xi
\\
\Conj{ \xi }
\end{pmatrix}
,
\begin{pmatrix}
z'
\\
\Conj{z'}
\end{pmatrix}
\right\rangle
&
=
\frac{1}{4}
\left\langle
\begin{pmatrix}
\xi
\\
\Conj{ \xi }
\end{pmatrix}
,
\begin{pmatrix}
1 & 1
\\
1 & -1
\end{pmatrix}
\begin{pmatrix}
1 & 1
\\
1 & -1
\end{pmatrix}
\begin{pmatrix}
z'
\\
\Conj{z'}
\end{pmatrix}
\right\rangle
\\
&
=
\frac{1}{4}
\left\langle
\begin{pmatrix}
1 & 1
\\
1 & -1
\end{pmatrix}
\begin{pmatrix}
\xi
\\
\Conj{ \xi }
\end{pmatrix}
,
\begin{pmatrix}
1 & 1
\\
1 & -1
\end{pmatrix}
\begin{pmatrix}
z'
\\
\Conj{z'}
\end{pmatrix}
\right\rangle
\\
&
=
\frac{1}{2}
\left\langle
\begin{pmatrix}
\xi + \Conj{\xi}
\\
\xi - \Conj{\xi}
\end{pmatrix}
,
\frac{1}{2}
\begin{pmatrix}
z' + \Conj{z'}
\\
z' - \Conj{z'}
\end{pmatrix}
\right\rangle
\\
&
=
\frac{1}{2}
\left\langle
\begin{pmatrix}
\xi + \Conj{\xi}
\\
\xi - \Conj{\xi}
\end{pmatrix}
,
\begin{pmatrix}
x'
\\
i y'
\end{pmatrix}
\right\rangle
\\
&=
\frac{1}{2}
\left\langle
\begin{pmatrix}
\xi + \Conj{\xi}
\\
i(\Conj{\xi}-\xi)
\end{pmatrix}
,
\begin{pmatrix}
x'
\\
y'
\end{pmatrix}
\right\rangle.
\end{align}
\end{subequations}
\Cref{equation-identifying-xbar-and-ybar} gives $\overline{x} = (\xi + \Conj{\xi})/2$ and $\overline{y} = i(\Conj{\xi} - \xi)/2$.
We map those two values to $\overline{z}$,
\begin{align}
\overline{z} 
= \overline{x} + i \overline{y}
= \frac{1}{2}\left( \xi + \Conj{\xi} + i^2 (\Conj{\xi} - \xi)\right)
= \xi
\end{align}
and see that $\overline{z} = \xi$ is the latent VJP.
This insight is valuable because it means that we can skip computing $\overline{x}$ and $\overline{y}$ and instead apply the identity 
\begin{align}
\label{equation-latent-vector-Jacobian-product}
\langle \overline{f}, \hat\partial f(z)(z') \rangle
+ \langle \Conj{ \overline{f} }, \hat\partial \Conj{ f (z) } (z') \rangle
=
\langle \overline{z}, z' \rangle
+
\langle \Conj{\overline{z}}, \Conj{z'} \rangle
,\end{align}
compute $\overline{z}$ from $\overline{f}$ and the Wirtinger derivatives of $f$, and read off the latent VJP $\overline{z}$. 
Propagating latent VJPs according to \Cref{equation-latent-vector-Jacobian-product} is reverse-mode differentiation with complex numbers.
For example, for $f(z) = \Conj{z}^\top A z$, we have the latent JVPs,
\begin{subequations}
\begin{align}
\hat\partial f(z)(z') &= z^\top A^\top \Conj{z'} + \Conj{z}^\top A z',
\\
\hat\partial \Conj{f}(z)(z') &= \Conj{z^\top A^\top} z'  + z^\top \Conj{A} \Conj{z'},
\end{align}
\end{subequations}
which leads to the transposition rule (in Wirtinger form),
\begin{subequations}
\begin{align}
&
\langle
	\overline{f},
	z^\top A^\top \Conj{z'} + \Conj{z}^\top A z'
\rangle
+
\langle
	\Conj{\overline{f}},
 \Conj{z^\top A^\top} z'  + z^\top \Conj{A} \Conj{z'}
\rangle
\\
&\quad\quad
=
\langle
	\Conj{A z} \overline{f}
	+ A^\top \Conj{z} \Conj{\overline{f}}
	,
	\Conj{z'}
\rangle
+
\langle
	\Conj{A}^\top z \overline{f}
	+
	A z \Conj{\overline{f}}
	,
	z'
\rangle.
\end{align}
\end{subequations}
Matching this expression to \Cref{equation-latent-vector-Jacobian-product}, which means collecting all terms in the same inner product as $z'$ and discarding all other terms, implies the latent VJP of $f(z) = \Conj{z}^\top A z$,
\begin{align}
\label{equation-latent-vjp-inner-product}
\overline{z} =
\Conj{A}^\top z \overline{f} + A z\, \Conj{ \overline{f} }
.
\end{align}
If we adopted a different gradient convention, the roles of $\overline{f}$ and $\Conj{\overline{f}}$ as well as the roles of $\overline{z}$ and $\Conj{ \overline{z}}$ would change.
\Cref{figure-reverse-mode-differentiation-complex} turns \Cref{equation-latent-vjp-inner-product} into \jax{} code.
\begin{figure}[t]
\par\bigskip
\begin{center}
\scalebox{0.9}{
\begin{tabular}{ p{0.9\linewidth}}
\begin{lstlisting}[language=Python]
import jax
import jax.numpy as jnp

key = jax.random.PRNGKey(1)
key1, key2, key3 = jax.random.split(key, num=3)

A = jax.random.normal(key1, (3, 3), dtype=complex)
x = jax.random.normal(key2, (3,), dtype=complex)
df = jax.random.normal(key3, (), dtype=complex)


def fun(v):
    print("Calling the function, not the VJP...")
    return jnp.dot(v.conj(), A @ v)


_, vjp_ad = jax.vjp(fun, x)  # Calls fun, not VJP
(dx_ad,) = vjp_ad(df)


def fun_fwd(v):
    print("Calling the forward pass...")
    return jnp.dot(v, A @ v), {"v": v}  # Cache some values


def fun_rev(cache, fbar):
    print("Calling the backward pass...")
    fbar = fbar.conj()  # switch convention
    zbar = (A.conj().T * fbar + A * fbar.conj()) @ cache["v"]
    return (zbar.conj(),)  # switch convention


fun = jax.custom_vjp(fun)
fun.defvjp(fun_fwd, fun_rev)
_, vjp_custom = jax.vjp(fun, x)  # Calls VJP, not fun
(dx_custom,) = vjp_custom(df)

assert jnp.allclose(dx_ad, dx_custom)
\end{lstlisting}
\end{tabular}
}
\caption{Complex reverse-mode differentiation in \jax{}.
Note the gradient convention switch.
}
\label{figure-reverse-mode-differentiation-complex}
\end{center}
\end{figure}

\paragraph{Consistency with holomorphicity.}
If the function $f$ is holomorphic, $\partial_\Conj{z}f = 0$ holds (which is another formulation of the Cauchy--Riemann equations \citep{hunger2007introduction}).
Thus, the latent VJP reduces to the VJP since the latent JVP matches the (true) JVP and both terms of the summand in \Cref{equation-latent-vector-Jacobian-product} can be matched independently.
In this case, \Cref{equation-latent-vector-Jacobian-product} simplifies to finding $\overline{z}$ according to
\begin{align}
\langle \overline{f}, \hat\partial f(z)(z') \rangle
=
\langle \overline{z}, {z'} \rangle.
\end{align}
Thus, if $f$ is holomorphic, the latent gradient becomes the (actual) gradient \citep{kreutz2009complex}.
However, if the function is not holomorphic, both terms in \Cref{equation-latent-vector-Jacobian-product} cannot be matched independently because, for example, $\hat\partial f(z)(z')$ depends on both $z'$ and  $\Conj{z'}$.
If the function maps to the real values, $\overline{z}$ points to the steepest ascent of $f$ \citep[Corollary 5.0.2]{hunger2007introduction}. 
If the gradient convention changes, $\Conj{\overline{z}}$ points in the same direction instead of $\overline{z}$; see this\footnote{\scriptsize \texttt{https://jax.readthedocs.io/en/latest/notebooks/autodiff\_cookbook.html\#complex-numbers-and-differentiation}} section in \jax{}'s documentation.

\section{Shortcomings of latent VJPs and latent JVPs}
\label{section-shortcomings}

Complex-valued automatic differentiation via latent JVPs,  latent VJPs, and Wirtinger calculus isn't without issues, especially if one expects similar behaviour to real-valued differentiation.
The main problem is unexpected behaviour when computing gradients of a function $f$ that is not differentiable.
In real arithmetic, the Jacobian matrix of $f$ would contain undefined values (usually ``NaN'' or $\infty$), which is a clear indicator that the derivative of a non-differentiable function has been attempted.
However, in complex arithmetic, latent JVPs and latent VJPs remain well-defined, even if a function is not differentiable.
This well-definedness may sound like an advantage, but it is actually a disadvantage because code for assembling a gradient returns an array without ``NaN'' or $\infty$ but with reasonably-looking complex values.
However, this gradient isn't the true gradient.
This silent failure means that a user must know whether a computer program is holomorphic to use complex automatic differentiation -- a difficult task.
Nonetheless, latent VJPs offer some relief because even if the (latent) gradient returned by the differentiation software may not be the actual gradient, it still points into the direction of steepest ascent and can be used for optimisation \citep[Appendix A]{fischer2005precoding}.

\section{Conclusion}
\label{section-conclusion}

This tutorial discussed Jacobian-vector and vector-Jacobian products in complex arithmetic, which implement forward- and reverse-mode automatic differentiation.
The main challenge with complex-valued automatic differentiation is that one can't assume that all functions are holomorphic, which necessitates a new kind of JVP and VJP.
The critical step for this new JVP/VJP is to regard a complex function as a function between higher-dimensional (``latent'') real-valued functions, implement JVPs and VJPs there, and map the results back into the complex numbers.
Wirtinger calculus changes the basis of $\Rbb^2$ to simplify this procedure, leading to simpler expressions for what we refer to as latent JVPs and VJPs.
If a function is holomorphic, these latent JVPs and VJPs reduce to the true JVPs and VJPs, leading to well-defined gradients and derivatives.

\section*{Acknowledgements}
This work was supported by a research grant (42062) from VILLUM FONDEN. The work was partly
funded by the Novo Nordisk Foundation through the Center for Basic Machine Learning Research in
Life Science (NNF20OC0062606).
This project received funding from the European Research Council (ERC) under the European Union’s Horizon programme (grant agreement 101125993).

I thank Stas Syrota and Hrittik Roy for helpful feedback on the manuscript. I am grateful to Simon Koop for correcting typographic errors in \Cref{equation-latent-vector-Jacobian-product,equation-definition-xi}, and to Emiliano Godinez for doing the same for \Cref{equation-wirtinger-derivatives}.

\bibliography{bibfile.bib}

\end{document}